\documentclass[%
reprint,
amsmath,amssymb,
pre,
]{revtex4-1}
\usepackage{bbold}
\usepackage{graphicx}% Include figure files
\usepackage{bm}% bold math
\usepackage{hyperref}% add hypertext capabilities
\usepackage[mathlines]{lineno}% Enable numbering of text and display math
\newcommand{\fig}[1]{(Fig. \ref{#1})}
\newcommand{\eq}[1]{(Eq. \ref{#1})}
\newcommand{\tab}[1]{(Tab. \ref{#1})}
\begin{document}
\title{Mathematics and Morphogenesis of the City: A Geometrical Approach}
\author{Thomas Courtat$^{1,2}$, Catherine Gloaguen$^1$, Stephane Douady$^2$}
\email{thomas.courtat , catherine.gloaguen (at) orange-ftgroup.com; douady (at) lps.ens.fr}
\affiliation{$^1$Orange Labs, 38-40, rue du G\'en\'eral Leclerc, 92794 Issy-les-Moulineaux, France \\ $^2$Laboratoire Mati\`ere et Syst\`emes Complexes (MSC), UMR CNRS
-Universit\'e Paris Diderot CC 7056, 10 rue Alice Domon et L\'eonie Duquet, 75205 Paris Cedex 13}
\date{\today}
\begin{abstract}
Cities are living organisms. They are out of equilibrium, open systems that never stop developing and sometimes die. The local geography can be compared to a shell constraining its development. In brief, a city's current layout is a step in a running morphogenesis process.
Thus cities display a huge diversity of shapes and none of traditional models from random graphs, complex networks theory or stochastic geometry takes into account geometrical, functional and dynamical aspects of a city in the same framework.
We present here a global mathematical model dedicated to cities that permits describing, manipulating and explaining cities' overall shape and layout of their street systems.
This street-based framework conciliates the topological and geometrical sides of the problem. From the static analysis of several French towns (topology of first and second order, anisotropy, streets scaling) we make the hypothesis that the development of a city follows a logic of division / extension of space. 
We propose a dynamical model that mimics this logic and which from simple general rules and a few parameters succeeds in generating a large diversity of cities and in reproducing the general features the static analysis has pointed out.
\end{abstract}
\maketitle
\section{Introduction \label{introduction}}
The city is a living structure: it is an open system, always in motion. A city is born, develops, heals over injuries (war damages...) and sometimes dies in part or totally. Its development responds to internal and external constraints in such a way that local geography acts as a shell that sculpts it. As general living systems, cities exhibit a huge range of diversity both on their overall shape (that can be circular, sprawling, linear or even fractal) and on the appearance of their street systems (regular, organic, tree-like). Such a diversity can also be observed at the level of a single city that has not developed homogeneously.
\\
We seek out to show that behind this diversity stands a single principle: a city develops within a logic of division / extension of space. The intra and inner diversity of shapes can be seen as a variation in a coherent global phenomenon. 
Our approach is street-based: we consider an infinitesimal piece of street as the elementary component of a city and bet it contains important information. The dual approach is to consider the build-up area (buildings, parks...) as the unit of formation. 
Several points of view have been used in the past to model cities: cellular automata, multi-agent systems, fractals, stochastic geometry, L-systems and graph theory leading to complex systems' theory. 
\\
Cellular automata and multi-agents systems have widely and successfully been used to simulate the dynamics of populations and of land use \cite{Sembolonia2004}. The fractal description of cities \cite{Frankhauser1998} gathers these simulations into a theory and point out the advantages of a fractal shaped city from the point of view of the build-up area. Nonetheless the basis of these models is either a discrete field or the map of a given city. They explain the global differentiation of space when the street network is known or ignored. 
\\
Since the famous "Bridges of Konigsberg" problem by Euler, one is tempted to describe a city as a graph with streets as edges and their intersections as vertices. This provides a relational representation of the city \cite{Blanchard2009}. One of the difficulties then is the particular embedding of these graphs that make random graphs unable to stick to a city's map representation. 
\\
Stochastic geometry gets around this problem by considering stationary tessellations (Poisson Vorono\"\i, Poisson Delaunay or Poisson Line Tessellation, their superpositions and iterations) that are geometrical objects embedded in a compact subset of the space and deduces geometrical random graphs from their induced topology \cite{Gloaguen2006}. L-systems with procedural programming make a map evolve from local coherence rules and input data that incorporate global constraints \cite{Parish2001, Chen2008}. 
The stochastic geometry approach gets good results at analyzing optimization problems on street networks and L-system are successfully used in graphics but they do not explain the underlying phenomena at work to determine the appearance of a city. 
\\
For a few years a complex networks based study has been adopted to describe cities \cite{Porta2006, Buhl2006}. But the main conclusion is that a city behaves neither like a classical scale-free network nor a small-world essentially because of its spatial embedding \cite{Boccalettia2006}. 
Cities then clearly need a dedicated mathematical framework taking into account both the topology and, what is new, the geometry of a street network. The scope of this article is a street-based approach of cities that allows analyzing, manipulating and explaining cities' morphogenesis. 
In (\ref{space}) we define a mathematical formalism to handle with cities both on a relational and a geometrical way. The very stuff of the geometry being that street segments are lined up into sets called streets we model via a hypergraph additional structure.
\\
(\ref{parameters}) presents several description measurements (topology of first and second order, anisotropy and streets scaling) to obtain quantitative comparison elements between cities and exhibit some features of the global mechanic of cities system (organic shape, log scaling of length and small topological radius). These features call for the modelling of the city as a process of division and extension of space. We use 10 French towns and their centers without having to restrict them to a square window to illustrate this.
\\
Eventually we present a morphogenetic model of the city (\ref{model}) and its simulation (\ref{simulation}) that implements the idea of space division / extension. 
This model reproduces the general features pointed out in (\ref{parameters}) and the variation of a few normalized parameters allows recovering a large range of diversity. The model is expressed in terms as relevant for mathematicians or physicists as for town-planners or social scientists. No sampling of space is required for the simulation.
\section{Cities' space \label{space}}
Since the famous resolution of the "Bridges of Konigsberg" problem by Euler \cite{Blanchard2009} one is tempted to look at cities with a formal and relational point of view: a city is a graph whose edges are streets and vertices are their intersections. 
\\
Nevertheless this approach does not take into account the physical constraints (of being a functional object on the plan) that are exerted on cities. Furthermore, it enhances a fundamental disjunction between intersections - that would be objects of interest - and street segments that would simply bind them. Under those street segments / edges lies a characteristic geometry. Each point of this geometry should be seen as an object in relation with other similar objects. This paragraph aims at introducing a terminological frame that permits manipulating cities as "continuous graphs" embedded in a two dimensional Euclidian space. The vocabulary used in this article is freely adapted from general graph theory \cite{Gross2004} to respond our to specific needs. 
\\ 
When importing a map (via a .MIF file), the raw data is coded into a list of polylines. In fact these polylines come from the sampling of curved streets that are difficult to represent in a computer. We consider it is possible to transform a \textit{geometrical graph} into a \textit{straight graph} arbitrary close to it.
The important paradigm of degree two points comes up. To \textit{rectify} the map we have added degree 2 vertices but we want to consider a version the same object. In short, a segment with its two extremities and the same segment with its extremities plus its midpoint should be seen as the same entity. We define a measure on a geometrical graph which allows to see that graph as a "continuum" and to consider each point of its geometry in the same time.
The data rather represents \textit{street segments} that actual \textit{streets}. The particularity of a city's geometry is that its street segments are coherently arranged into disjoint geometrical sets: the streets. We try to record the notion of streets with a \textit{hypergraph} additional structure. This provides a multi-scale representation of the city.
\subsection{Graphs and planar graphs}
\begin{figure}[!h]
\begin{center}
\includegraphics[scale=1]{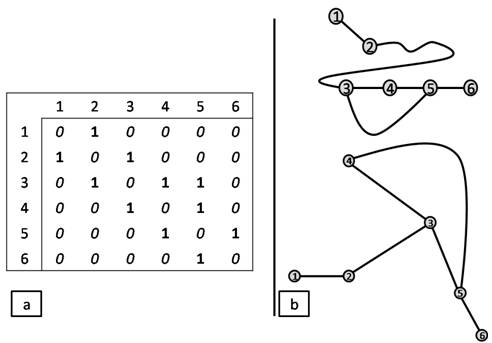}
\caption{ The representation of a graph by its symmetric adjacency matrix (a). This graph is planar: it admits at least two geometrical graphs as drawings (b). }
\label{graphToGeometrical}
\end{center}
\end{figure}

Let $S$ be a set, $V$ (for vertices) a finite subset of $S$, $E$ (for edges) a symmetric part of $V \times V$ then $G = (V,E)$ is said to be a (undirected) graph. 
\\
A drawing of $G$ is an injective function from $V$ to $\mathbb{R}^2$ and from $E$ to the set of continuous paths such that the image of an edge has for limits the images of the vertices it binds and does not pass through images of vertices it does not bind. An edge-crossing is the intersection of the images of two edges outside the image of $V$.
\\
If there exists a drawing without any edge-crossing, the graph is said to be planar \fig{graphToGeometrical}. The first characteristic of city graphs is their planarity. 
\subsection{Geometrical graphs} 
\begin{figure}[!h]
\begin{center}
\includegraphics[scale=1]{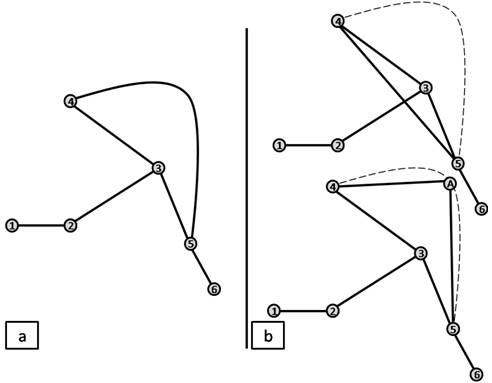}
\caption{ A geometrical graph (a) and two of its rectifications (b). The upper one is not straight because two edges intersect outside of the vertices set.}
\label{graphRectification}
\end{center}
\end{figure}

A geometrical graph can be seen as a particular drawing of a planar graph. 
\\
Let the available space $\mathbb{A}$ be a connected and compact subset of $\mathbb{R}^2$, $V$ a finite subset of $\mathbb{A}$ and $E$ a set of almost everywhere derivable paths included in $\mathbb{A}$ from one element of $V$ to another that do not intersect outside of $V$. Then $G=(V,E)$ is an element of the space of geometrical graphs $\mathcal{G}_g(\mathbb{A})$. If $E$ is restricted to straight segments ($ G \in \mathcal{G}_s(\mathbb{A})$) , $G$ is a straight graph.
To a geometrical graph $G$, one associates $\pi_G$, the subset of $\mathbb{A}$ defined by : 
\begin{equation} 
\pi_G = \{ x \in \mathbb{A}, \ \exists e \in E, \ x \in e \}
\end{equation} 
$\pi_G$ is compact so we can provide $\mathcal{G}_g(\mathbb{A})$ with an Hausdorff distance : 
\begin{equation}
d_H(g_1 \, || \, g_2) = \max_{x \in \pi_{g_1}} \, \min_{y \in \pi_{g_2}} \ ||x-y||
\end{equation}
A drawing $G' = (V',E')$ is a rectification of the geometrical graph $G = (V,E)$ if $V \subset V' \subset \pi_G$ and if each element of $E'$ is a segment. $G'$ is not necessarily a planar straight graph since edges can possibly intersect outside of vertices \fig{graphRectification}. The idea is that one should be able to add to a geometrical graph as many vertices of degree two as he wishes and still consider the same mathematical object. In this article we will admit that Pr1 and Pr2 hold for a large enough class of geometrical graphs: 
\begin{description}
\item[Pr1]: Every geometrical graph admits a planar rectification 
\item[Pr2]: Every geometrical graph is the limit of a sequence of straight graphs 
\end{description}
To get (Pr1), one has to sample an original graph with enough additional vertices of degree 2 and a small enough edges length. (Pr2) states that it is possible to approximate any geometrical graph by a straight graph arbitrary close to it in the sense of $d_H$. This is of practical importance as it allows to work only with straight planar graphs, while original city's maps can have curved edges.
\subsection{Measure}
As a compact part of $\mathbb{A}$, $\pi_G$ is a polish space (complete and separable) on which one can define a borelian measure $\mu_G$. For instance $\int_G d\mu_G$ is the total length of edges in $G$ or in (\ref{anisotropy}) we define the angular density of a graph $\Psi(.)$ with $\mu_G(.)$.
This measure respects: 
\begin{equation}
\int f(g) \, d\mu_G = \sum_{e \in E}| \int_e f(x) \, d\mu_1(x)| 
\end{equation}
Where $\mu_i$ is the $i$-dimensional borelian measure, $f$ a positive function and $\int_e$ the integral along the path $e$. If $f$ is a continuous function defined on $\mathbb{R}^2$ we have : 
\begin{equation}
\int f(g) \, d\mu_G = \lim_{\epsilon \to 0} \frac{1}{2.\epsilon}\int f(x,y). (\pi_G \oplus B_{\epsilon}) \, d\mu_2(x,y)
\end{equation}
where $\oplus$ is the Minkowski addition and $B_{\epsilon}$ the closed Euclidian ball of radius $\epsilon$ : $x \oplus B_{\epsilon} = \{ \ y \in \mathbb{R}^2 \ ||x-y|| \leq \epsilon \ \}$. So $\mu_G$ is a measure between $\mu_1$ and $\mu_2$ that allows to make quantitative measurements on the whole geometrical graph.
\subsection{Hypergraph structure}
If $H$ is an equivalence relationship on $E$ then $((V,E),H)$ is said to be an hypergraph. 
\\
Let $(V,E)$ a graph and $R$ a reflexive relationship on $E^2$. Then the relationship $\hat{R}$ defined by : 
\begin{multline} 
e_1 \, \hat{R} \, e_2 \ \text{iif} \ \exists \: \alpha_1=e_1, \, \alpha_2,\, ... \, , \, \alpha_n = e_2 \in E \, | \\ \alpha_1 \ R \ \alpha_2, \, \alpha_2 \ R \ \alpha_3, .... , \alpha_{n-1} \ R \ \alpha_n \end{multline} 
is an equivalence relationship. From this, one can consider $R_\theta$ : 
\begin{equation} e_1 \ R_\theta \ e_2 \quad \text{iif} \quad (e_1 \star_2 e_2) \vee ((e_1 \star e_2) \, \wedge \, (|\measuredangle(e_1,e_2)- \pi | \leq \theta)) \end{equation}
where $e_1 \star e_2$ means that $e_1$ and $e_2$ intersect, and $e_1 \star_2 e_2$ that $e_1$ and $e_2$ intersect in a vertex of degree 2. $\measuredangle(e_1,e_2)$ stands for the angle between $e_1$ and $e_2$ oriented with the same origin.
\\
For instance think of considering the map of a city and the relationship "these two edges are pieces of the same street". This $ R_\theta$ allows recovering the notion of "streets" even if input data do not contain such labels. The algorithm labeling streets segments with a street number does not depend on its starting point and is fast to run. The price to pay is that some special cases as forks of two segments making a very small angle with a third one will be considered as a single street. 
\begin{figure}[!h]
\begin{center}
\includegraphics[scale=1]{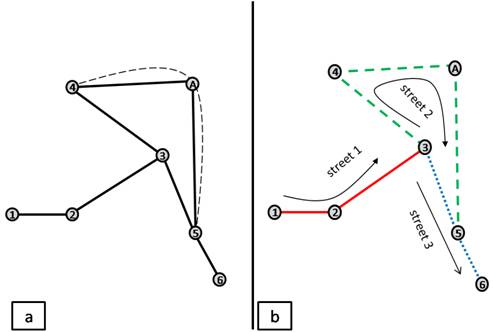}
\caption{A straight graph (a) and its hypergraph structure (b) deduced from $R_{\pi/20}$. Viewed as a city's map, this graph contains 7 streets segments but 3 streets. }
\label{graphToHyperGraph}
\end{center}
\end{figure}
This over structure is essential as it gives a way to analyze overall structures of planar graphs, and in particular of cities.  
\subsection{Citiy graphs}
We define pragmatically the set of city graphs $\mathcal{G}_C$ as the subset of $\mathcal{G}_S$ that represents an existing city or a city that could have existed. We restrict this definition to $\mathcal{G}_s$ since a lot of cities' streets are straight and even if it is not the case there exists a straight approximation as accurate as we want.
\\
In the following, we will write $ C =(V,E)$ a city and canonically provide $C$ with an hypergraph structure from a relationship $R_{\theta}$. For the sake of simplicity we keep the same notation to designate the hypergraph : $ C =((V,E),H)$. Its borelian measure is written $\mu_C$. 
\\
The relational aspect inherited from a simple graph structure allows us to define the set of faces $F$. Euler's equality is respected so that: $\sharp V - \sharp E + \sharp F =1 $. To each edge $e$ we associate the set $V(e)$ of its extremities in $V$ and to each point $v \in \pi_C$ we associate $E(v)$ the set of edges that pass through it. $N(v)$ is the degree of a vertex. 
\\
We partition $V = V_1 \cup V_2 \cup V_+ $ where $V_1$ contains all vertices of degree 1, $V_2$ of degree 2 and $V_+$ of higher degrees. Vertices in $V_1$, $V_2$ and $V_+$ are respectively terminations, junctions and intersections. Elements of $V_2$ will be seen has sampling artifacts used to fit the curved geometry of the city. 
\\
Elements of $E$ are called streets segments, those of $H$ streets. 
\section{ City space features \label{parameters} }
Due to their spatial constraints, real geometrical graphs do not behave as classical complex networks (small-worlds or scale-free networks) \cite{Boccalettia2006}. 
Real cities display structural, geometrical and functional features. 
For instance a real city aims both at lodging its inhabitants and at providing them with an efficient access to geographical and human resources. 
These constraints logically affect the structure of a city graph. 
The purpose of this part is to define some mathematical tools that will quantitatively measure structural differences between city graphs. Classical measures from complex networks theory (efficiency, robustness, centrality, degree correlation) have been investigated in \cite{Crucitti2006, Jiang2004, Cardillo2006}. The measures presented below are dedicated to cities.
Each one is illustrated with the city of Amiens in France: the whole city and its center \fig{amiensTopologicalMap}. The main properties are recapitulated in (\ref{synthesis}) for 10 French towns and their centers.   
\\
We have imported vector maps in a calculus framework and successively: rectified them by taking care of conserving the planarity and angles at the intersections and used $R_{\pi/10}$ to obtain an hypergraph structure.
\subsection{First order topology}
\begin{figure}
\begin{center}
\includegraphics[scale=1]{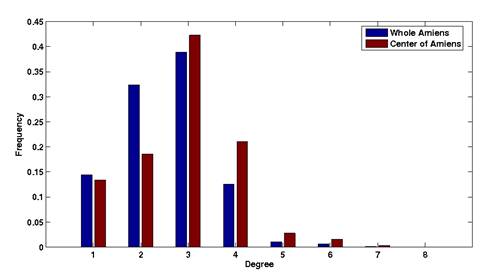}
\caption{ The histograms of degrees' distribution in Amiens (whole city and center). We observe that both distributions are peaked in 3 even in the more regular city centre, where the number of connections 4 is still half. The large number of 2 comes from straightening, especially in the suburb curved streets and should not be taken into account. 
}
\label{amiensDistDegreeVertices}
\end{center}
\end{figure}

Let $C=((V,E),H))$ be a city and $ N(k) = \sharp\{v \in C, N(v) =k \}$ be the number of vertices of degree $k$ in $C$ and $\bar{N}(k)= N(k) / \sum N(i)$. The set $V_2$ (junctions) should not be taken into account since it only represents sampling artifacts to preserve the shape of streets. In \cite{Buhl2006} the histogram of $N$ is studied by means of an exponential tail of distribution. Nonetheless, this distribution \fig{amiensDistDegreeVertices} is very peaked in 3 or in 4. It is sufficient to describe the histogram $N$ by the organic ratio: 
\begin{equation}
r_N = \frac{N(1)+N(3)}{ \sum_{j \neq 2} N(j)}
\end{equation}
which allows to discriminate quickly whether the city had been planned ($r_N \simeq 0$ in the limit case) or not. Indeed a planned city is filled with a homotopy of a rectangular grid (only $ \bar{N}(4) \neq 0$). This is clearly useful to settle buildings but also sticks to human perception of space since we have the intuition of left - right / front - behind. In unplanned cities (organic will be the dedicated word in \ref{model} to emphasize the comparison with living systems) i.e. created by the interaction between non concerted settlements, there is little probability for streets segments to be coherent and thus $r_N \simeq 1$.
\\
Following \cite{Cardillo2006} we characterize the topology of a city by its " meshedness coefficient". It's easy to count $v$ and $e$ then $f$ is deduced from Euler's formula. 
Given $V$, the maximum number $2.s-5$ of faces is obtained by the Greedy Triangulation algorithm \cite{Cardillo2006}. So the quantity $ M= (e-v+1)/(2.v-5) $ equals to 0 if the city is a tree and is close to 1 if it is a highly connected graph. For real cities \cite{Jiang2004} $M$ typically ranges between $0.08$ and $0.35$. 
\\
To be coherent with our preceding remark we should not take junctions into account. It has no incidence on the numerator but we have to change $2.v-5$ into $2.v.(1-\bar{N}(2))-5$
\begin{equation}
M_3 = \frac{e-v+1}{2.v.(1-\bar{N}(2))-5}
\end{equation}
$M_3$ is quite small because of the general lack of triangles in the topology of a city. As a trapezoid contains two triangles, we rescale it in $M_4=2M_3$ whose maximum is hit when the considered city contains the maximal number of trapezoids. Amiens appears as an "average" organic city, with a meshedness coefficient of $0.41$ ($0.54$ when restricted to the center) and $r_N=0.79$ ($0.68$ in the center).
\subsection{Second-order topology}
The streets $H$ induce a particular topology. In \cite{Jiang2004} its study is referred as the "dual approach". We use here the expression "second order topology" since it is a derived topology. Moreover, as faces are the dual of vertices, the dual of a space is a space containing the same information. We prefer to keep back the word "duality" to express this idea: " the mass of a city (buildings, houses, parks) is the dual of the street system" which could refer to the work of \cite{Frankhauser1998, Sembolonia2004}. 
\\
Let us call topological distance of $C$ the function $d^{topo}_C : H \times H \longrightarrow \mathbb{N}$ that satisfies:
\begin{equation} 
\left\{ 
\begin{array}{llllc}
d^{topo}_C(h_1 , h_2 ) &= &0 & \text{if} \, h_1 = h_2 \\
d^{topo}_C(h_1 , h_2) &=& \underset{h \in H, h \cap h_2 \neq \emptyset }{\min} {d^{topo}_C(h, h_1)}+1 & \text{otherwise} 
\end{array}
\right.
\end{equation} 
The topological distance counts the number of times one needs to turn to go from a street to another one. 
The topological average distance of a street $h_0$ is :
\begin{equation} 
\bar{d}^{topo}_C(h_0) = \frac{1}{\sharp H}\sum_{h \in H} d^{topo}(h, h_0)
\end{equation} 
This formula defines a new centrality measure on the map similarly to those studied in \cite{Crucitti2006}.
A street that minimizes $\bar{d}^{topo}_C$ is then called a center.
One drawback of this wholly topological definition is that, since topological distances are integers, several streets  can be defined simultaneously as central streets.
Common sense would then be to take all of these streets as simultaneous central streets, and to calculate the distance of any street as the minimum of the distance to any of these streets.
Another way is to weight by the streets length:
\begin{equation} 
\bar{d}^{topo \ leng}_C(h_0) = \frac{1}{\mu_C(C)}\sum_{h \in H} ||h||.d^{topo}(h, h_0)
\end{equation} 
($||h||$ is the length of the street $h$). From this a unique center $h_c$ is defined if the city is not too regular. The topological radius of the city is then defined by $r^{topo}_C = \max d^{topo}_C(h, h_c) $ and its diameter by : 
\begin{equation} 
\text{diam}^{topo}_C = \underset{h_1,h_2 \in H}{\max} \quad d^{topo}_C(h_1, h_2) 
\end{equation} 
\begin{figure*}
\begin{center}
\includegraphics[scale=1]{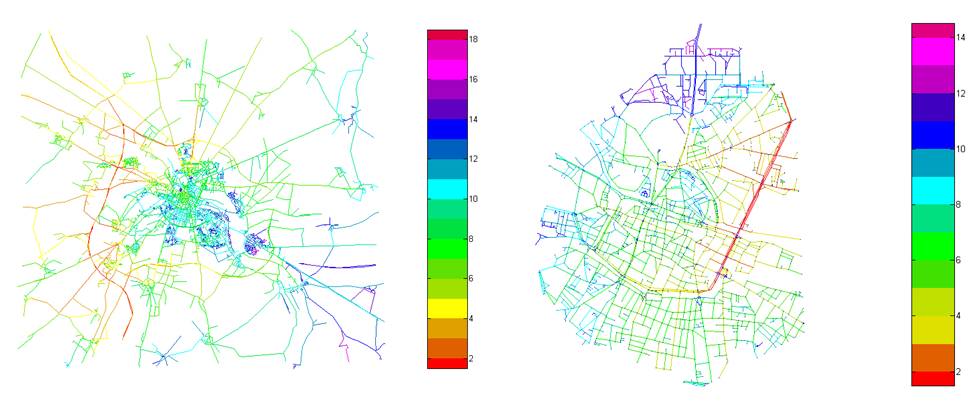}
\caption{(Color online). The "topological maps" of Amiens (left) and its center (right). In each map, the red street is the topological center and the color of each street refers to its distance to that center. The maximum distance to the center is the radius of the map (18 in the whole city and 16 in the center). It is striking that this radius increases slowly with the size of the considered street system. We can even infer that the construction of surrounding highway belt (the found center) is made precisely to keep the topological radius of the city small.}
\label{amiensTopologicalMap}
\end{center}
\end{figure*}
with $r_C^{topo} \leq diam_C^{topo} \leq 2r_C^{topo}$ that make these measures be equivalent. \fig{amiensTopologicalMap} plots in a color map the distance of each street to the topological center of Amiens that ends up to be a part of its highway-belt. This map gives a hierarchical vision of the space. There is no radial component of the increase of the topological distance: a scale of long streets serves the whole city, allowing the variation of the topological distance to be mainly local.
\\
Added to that the topological radius of the city grows very slowly with the size of the city (14 in the center of Amiens, 18 in the whole city that is eight times bigger).
\\
The topological efficiency $1/\bar{d}^{topo \ leng}_C(h_0)$ is defined for each street and we can consider it is defined on each point of the city, beeing constant almost everywhere in a street.It defines a new centrality \cite{Crucitti2006} on the network.
%\subsection{Compactness}
%The fractal theory of cities \cite{Frankhauser1998} shows that a city is most of the time not a compact object. This as much in the overall boundary of the city as in the spatial distribution of infrastructures. 
%\\ 
%We propose here an easy to calculate indicator that express in the same time the compactness (or fractality) of the city's shape and of its street system. 
%\\ 
%Let $\mathfrak{A}$ be the area of the convex hull of a city graph $C$. $\mu_C(C)$ is the total length of its street system. 
%\\
%The we imagine a city with a square convex hull of the same area that is to say with a side of length $\sqrt{\mathcal{A}}$ and a regular square lattice filling this hull with the same total length of streets. The area of mesh divided by the area of the hull is a number between $0$ and $1$ and one minus this quantity is a measurement of the idea of compactness in the city. Thus we define: 
%\begin{equation}
%\text{Comp}= 1-\frac{4\mathcal{A}}{(\mu_C(C)-2\sqrt{\mathcal{A}})^2}
%\end{equation}
\subsection{Anisotropy \label{anisotropy}}
\begin{figure}
\begin{center}
\includegraphics[scale=1]{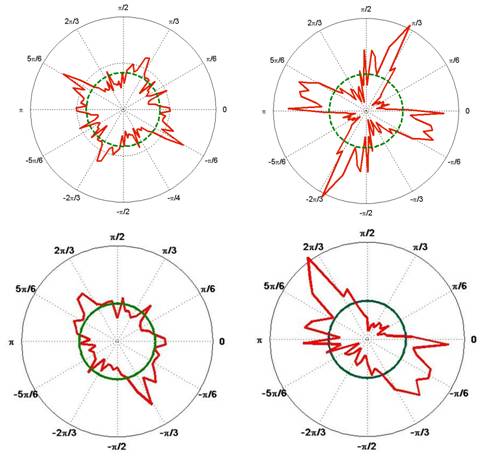}
\caption{Angular distribution of Amiens (top left : whole city - top right : center). We notice on both distributions a four-fold symmetry. The doubled angular distributions (bottom) look like ellipso\"{\i}ds which allows to define the anisotropy coefficient as the ratio of the eigenvalues of the inertia matrix: $Ani=0.42$ in the whole city and $Ani=0.71$ in the center.}
\label{AmiensAniso}
\end{center}
\end{figure}

In order to grant an efficient access to physical resources, the street system locally tends to be perpendicular to \textit{structuring elements} as for example rivers or older streets.
Then a city is not "isotropic". 
\\
Let $\vec{u}_0 \in \mathbb{R}^2$ be an arbitrary vector, taken as an angular origin. 
For angle $\alpha \in [0,\pi]$ , 
\begin{equation} 
\Psi^*(\alpha) = \mu_C \left( c \in C , \, \measuredangle (c,\vec{u}_0) \in [0, \alpha] \right)
\end{equation} 
where $ \measuredangle (.,.)$ is the angle measure between two vectors in $[0 , \pi]$. It is a measure of the "total length" of streets in $C$ that are oriented in directions $[0, \alpha]$; in the special case of straight graphs, the impact of each street-segment is proportional to its length. From this, we define the angular density by $ \Psi(\alpha) =\frac{d\Psi^*(\alpha)}{d\alpha}$ which representation describes the anisotropy of the city graph \fig{AmiensAniso}. We notice a fuzzy symmetry around the first bisectrix as a result of the streets local perpendicularity. For an isotropic city, the angular density $\Psi_I$ would be a continuous and uniform density $ \Psi_I(\alpha) = \frac{1}{\pi} $.
\\
It would be useful to sum up this angular density as a single normalized indicator. 
We looked for a bound distance measure between $\Psi$ and $\Psi_I$. Since the observed distribution $\Psi$ is discrete because of the limited number of streets segments, measures like $ \int |f-\frac{1}{\pi} |^n $ highly depends on the size of the bean chosen to estimate the integral. 
\\
Since the angle is defined modulo $\pi$, we can "fold" $\Psi$ : $\Psi^{(2)}(\theta)= \Psi(\theta)e^{i2\theta} - \int \Psi(u)e^{i2u}du $
The inertia matrix
$$ \left( \begin{array}{cc}
\int \mathfrak{Re}(\Psi^{(2)})^2& -\int \mathfrak{Re}(\Psi^{(2)})\mathfrak{Im}(\Psi^{(2)}) \\ 
-\int \mathfrak{Re}(\Psi^{(2)})\mathfrak{Im}(\Psi^{(2)}) & \int \mathfrak{Im}(\Psi^{(2)})^2
\end{array} \right) $$ 
is symetric and positive (from Cauchy-Schwartz's inequality) with two eigenvalues $\lambda_1 > \lambda_2$ such that 
\begin{equation}
Ani=1-\frac{\lambda_2}{\lambda_1}
\end{equation}
correctly defines an anisotropy coefficient. As for Amiens, its anisotropy coefficients varies from $0.42$ in the whole city to $0.71$ in the center.
\subsection{Street length}
\begin{center}
\begin{figure}
\includegraphics[scale =1]{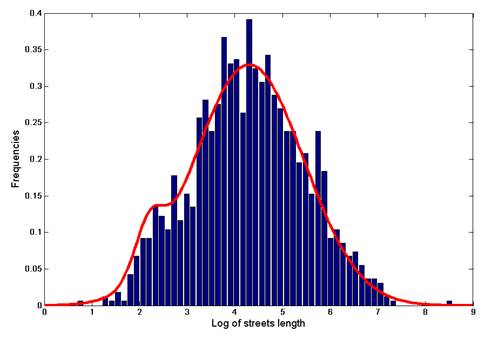}
\caption{The histogram represents the distribution of the logarithm of streets length in the center of Amiens. The red plot is the fitting of this histogram with a gaussian mixture whose parameters are $p_-=0.5$, $m_-=2.2$, $\sigma_-=0.3$; $m_+=4.3$, $\sigma_+=1.2$ . }
\label{AmiensCenterStreetsLength}
\end{figure}
\end{center}

The second order topology leads to think that a city organizes into a hierarchical way with long and no so long streets. Consequently we do not expect an exponential decay in the distribution of street lengths. 
In Amiens, the distribution of street lengthes $L$ \fig{AmiensCenterStreetsLength} is well-fitted by a mixture of Log-Normal laws: 
\begin{equation}
\log{L} \sim p_-.\mathcal{N}(m_-, \sigma_-) + (1-p_-).\mathcal{N}(m_+, \sigma_+) 
\end{equation}
with $m_-<m_+$. The identification of this model is performed with an Expectation Maximization algorithm.
\\ 
The log-scaling reveals that there is no evident length scale in a city, there is no pre-existent typical street. A city auto-scales, its dynamics is purely multiplicative and could be the result of new streets cutting through former blocks or extending space at the exterior of the city.
\\
As for the bimodality, a town-planning explanation is that several transportation modes follow each other through time and their superposition creates modes in the distribution of $L$. 

\subsection{Synthesis: the city as a space division process \label{synthesis}}
\begin{table*}
\begin{tabular}{|c||c|c|c|c|c|c|c||c|c|c|c|c|c|c||}
\hline & \multicolumn{7}{|c||}{Whole cities} & \multicolumn{7}{|c|}{Centers} \\ 
\hline City $C\downarrow$ & $r_N$& $M_4$ & $Ani$ & $\mu_C(C)$ & $\sharp H$ & $r^{topo}$ & $res$ & $r_N$ &$M_4$ & $Ani$ & $\mu_C(C)$ & $\sharp H$& $r^{topo}$ & $res$ \\ 
\hline Angoulême & 0.80&0.28 &0.46 &300 &1628 & 14 & 0.21 & 0.73 & 0.37 & 0.29 & 39 & 261 & 7 & 0.2 \\ 
\hline Avignon & 0.85 & 0.23 & 0.59 & 625 & 3348 & 13 & 0.12 &0.82 & 0.30 & 0.73 & 240 & 1607 & 9 & 0.12 \\ 
\hline Caen& 0.79& 0.29 &0.53 & 485 & 3045 & 11 & 0.08 &0.76 & 0.34& 0.67 & 128& 797& 9& 0.22 \\ 
\hline Carcassonne& 0.86 &0.20 &0.66 &483 &1997 &20 &0.17 & 0.72 & 0.38& 0.89& 66& 296& 6 & 0.42 \\ 
\hline Dijon& 0.75& 0.33 &0.39 &558 &2605 &14 &0.14 & 0.66& 0.42& 0.74& 149& 860 & 8 & 0.33 \\ 
\hline Grenoble& 0.74 & 0.32 & 0.61 &361 &1638 &10 &0.17 & 0.70& 0.40& 0.74 & 119 & 576 & 6 & 0.29 \\ 
\hline Lyon& 0.66 & 0.47 &0.53 &837 &3606 &15 &0.17 & 0.53 & 0.51 & 0.93 & 183 & 606 & 6 & 0.24 \\ 
\hline Rennes& 0.82 &0.26 &0.79 &625 &3538 &15 &0.20 & 0.77& 0.30& 0.80 & 192 & 1041 & 7 & 0.17 \\ 
\hline Rouen& 0.71 &0.38 &0.69 &348 &1770 &17 &0.19 & 0.66 & 0.43 & 0.85 & 141 & 788 & 7 & 0.13 \\ 
\hline Troyes& 0.81 &0.28 &0.87 &230 &1079 &9 &0.17 & 0.67 & 0.42 & 0.92 & 48 & 248& 6 & 0.41 \\ 
\hline 
\end{tabular} 
\caption{ Features of 10 French cities and their centers. $r_n$ is the organic coefficient, $M_4$ the meshedness coefficient, $Ani$ the anisotropy, $\mu_C(C)$ the total street length, $\sharp H$ the number of streets in the hypergraph generated wit $R_{\pi/10}$, $r^{topo}$ the topological radius of that hypergraph and $res$ the root mean square between the street length distribution and its bi-lognormal fitting.} 
\label{french}
\end{table*}
\tab{french} summarizes the main indicators presented through this section for ten French cities and their extracted centers.
For French towns the organic coefficient $r_N$ may seem surprisingly high, even for Lyon's center which is visually grid-like. The preponderance of degree 3 intersections speaks in favor of seeing a city as the result of a division process. The cutting of a block creates new intersections of degree $3$ even if the block was formed of degree $4$ vertices.
\\
The meshedness coefficient $M_4$ is strongly correlated to $r_N$. In fact it is a second order refinement of $r_N$. We will show that it describes the shape of city graphs when $r_N$ is constant equal to $1$ in \tab{tabMesh}.
\\
The anisotropy coefficient $Ani$ is the most discriminating indicator: it ranges from $0.29$ to $0.93$. It also points out that the street system within a city is not homogenous: for instance $Aniso(\text{Lyon}) =0.53$ and $Aniso(\text{Lyon Center}) =0.93$.
\\
The total street length $\mu_C(C)$ and the number of streets $\sharp H$ measure the size of a city. In fact the ratio between $\mu_C(C)$ and $\sharp H$ provides an indication on the straightness of the city which is geometrical information.
\\
The topological radius $r^{topo}$ is very small compared to a measure of the city size: $r^{topo} \ll \sharp H$. $r^{topo}$ may scale as the logarithm of $\sharp H$, like a small-world complex network. This scaling may also be the consequence of a division process and/or of an upper system of long streets that covers the whole city's stretch constraining the topological distance's variations to be mainly local.
However the topological radius is an indicator of the transportation performance of a given city: Carcassonne and Caen have almost the same total street length but the topological radius of Carcassonne is roughly twice Caen's. 
\\
$res$ the relative root mean square between the street length distributions and their bi-lognormal fitting is $ \lesssim 0.2$ with two exceptions (Carcassonne and Troyes's centers) around $0.4$. The lognormal scaling of street lengths is relevant in most of the cases. This also calls for a division (multiplicative) modelling of the city. But a division process's length distribution would only have a single tail.
\\
In the following part (\ref{model}) we present a morphogenesis model for the organic city consisting in the duality between division and extension of space. In (\ref{simulation}) we will show that this model reproduces the features we have presented above: a high organic coefficient, a small topological radius and a local variation of the topological distance plus a lognormal distribution for streets' lengths. 
\section{A streets based dynamical model \label{model}} 
This section presents a model of the growth and development of a town. The town is reduced to its streets and we build a dynamical model allowing adding street segments one after another. As in the previous parts, the spatial extension of the town and the geometry of the streets is of prim interest. 
As pointed out in \cite{Barthelemy2009}, a city is above all an out of equilibrium system, that is to say a dynamical system observed at a random time of its development. 
\\
Our model is first on three assumptions, two principles (installation and connection) and a few parameters. The whole giving a coherent and consistent vision of the problem. We aim at building a model that can reproduce several limit cases of urban growth but also point out continuity between them.
The principles and parameters we use are meaningful, expressed in an interface language that allows the mathematical and physical communities to exchange with town planners, architects and social scientists.
\\
This part develops the model in the quite general case of an organic development of the city on flat lands for which we can easily translate our assumptions into analytical procedures. 
\subsection{Hypothesis \label{hyp}}
As a dynamical system and a geometrical graph, we will see a city as a function $ C : \mathbb{R}_+ \longrightarrow \mathcal{G}_C \subset \mathcal{C}_d$ with $C(t) = \lbrace V(t), E(t) \rbrace $. Then we make the following postulates on the evolution of $C$ : 
\begin{description}
\item[P1] A city is the result of a sequence of operations occurring at increasing times $(t_i)_{i \in \mathbb{N}}$ such that : 
$$ C(t) = C(t_i) \quad \forall t \in [t_i, t_{i+1}[ $$
\item[P2] Infrastructures are conserved: 
$$ C(t_1) \subseteq C(t_2) \quad \text{if} \ t_1 \leq t_2$$ 
\item[P3] There exist two functions $P_t$ (price) and $V_t$ (potentiality) such that the city is a compromise between them: 
$$ C(t + \Delta_t ) = \underset{\begin{array}{c} 
c \subset C(t) \\ 
P_t(c-C(t)) \leq 0 
\end{array}}{\text{argmin}} V_t(C(t), c) $$ 
\end{description}
Functions $C$ and $P$ are not obvious to define. They should be in a "microscopic" point of view aggregation of economical parameters. We can avoid developing them if we observe a city's growth is determined by "macroscopic" insights:
\begin{description}
\item[Its planning] A city may be organic (the sum of local and independent phenomena: streets are added independently with no visibility on a global planning) or centralized (a global authority decides of the coherent and simultaneous addition of several streets on a large surface).
\item[Its construction] The capacity to add new elements to the map and to build new streets.
\item[Its organization] From a random settlement to a highly structured one. 
\item[Its sprawling] A city has to make a compromise between its inner development and outer growth.
\end{description}
We will consider here the case of organic growth. 
\subsection{Organic growth of a city}
The algorithm below simulates a city's growth within the individual settlement hypothesis. 
\\
Under this assumption, each settlement (generic term to designate a commercial infrastructure, a private individual...) is added at a given time and at a given location then connects to the existing infrastructures. 
\\
The main idea is that the city $C(t)$ induces a spatial potential field describing the attractiveness of any point of the available space. A new settlement (either an individual settler or a facility) has its own policy (\ref{modPara}) of choice with respect to this potential. After having chosen its location, it connects to the existing street system. This model explicitly decouples the problems of positioning and of connecting.
\\
In this section we will model how the geometry of the current city induces a potential field, how a new settlement can be connected to the city and at last how we can tune the behavior of settlements by a few parameters.
\subsubsection{Potential field}
\begin{center}
\begin{figure}
\includegraphics[scale=1]{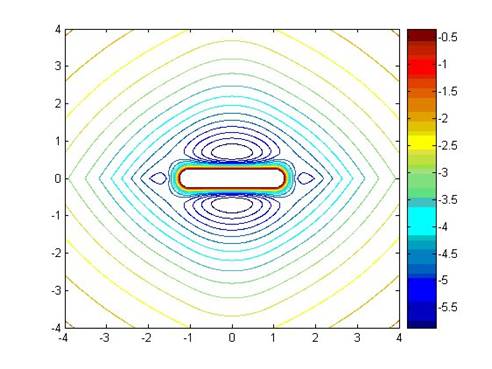}
\caption{(Color online). The level lines of the potential field for a city reduced to a single segment of length 1with $\lambda_0=1$ and $\beta=10$.}
\label{pot}
\end{figure}
\end{center}

For each point $x$ in the available space the potential $P_{C \to .}$ quantifies to what extent $x$ is a good choice to locate a new center. This potential should mimic the following ideas: 
\begin{itemize}
\item A large scale behavior such that the global attraction of a part of the city should be proportional to the global mass of infrastructures in place and slowly decrease with some distance $d$ : $ P_{C \to x}\propto -\frac{\int d\mu_C}{d^{\gamma}(x,C)}$ 
\item A very short scale behavior that forbids a new center to be located on existing infrastructures: $ P_{C \to x}= + \infty $
\item A medium scale deduced from the two previous ones, that should display some local minimum. 
\end{itemize}
Thus among several possible fields whe choose \fig{pot}:
\begin{equation}
P_{C \to x} = \left( \frac{\alpha}{d_{min}(x,C)} - \frac{\beta}{\sqrt{d_{min}(x,C)}} \right) \int \frac{d\mu_C}{\sqrt{d_{\bot}(c,x)}}
\label{epot}
\end{equation}
$d_{min}(x,C) = \min_{c \in C} \, (x,c)$ is used in \eq{epot} so that the rejecting zone is hard : there is a tube around the city where new settlements are impossible. The radius of this tube is $\lambda_0 = \left( \alpha/\beta \right)^2$.
\\ $d_{\bot}(x,c)$ is the $||.||_1$ norm in the local basis formed by the unitary tangent and the normal to $C$ in $c$. The use of such a distance simplifies integral calculus compared to Euclidian distance.
Towards $\infty$, $P_{C \to x} \sim \frac{\beta}{d(x,C)}$ and between those two extreme positions, interferences between streets segments produce local minima. 
To choose parameters $\alpha$ and $\beta$, one sets $\lambda_0$ the hard rejection radius and $\beta$ the long-range influence. The choice of $\beta$ influences the local geometry of the city but won't be discussed here. 
\\ 
Among all possible potentials, we picked one that fulfils the conditions we set and that allows an explicit calculus of the integral. Further discussions would address the choice of the used distance and the decay exponent $\gamma$.
\subsubsection{Connection} 
Once a settlement is added in a location $x$, it links to the existing network $C$. Not all connections are eligible. 
\\
From a point $x$ we define the visible set of points: 
\begin{equation}
V_{x | C}\left\lbrace x \in C, [c \ x] \cap C =\left\lbrace c \right\rbrace \right\rbrace 
\end{equation} 
And the optimal set of points from $x$ of a part $E$ of $C$: 
\begin{equation}
\dot{E}_x = \{e \in E \ \exists \epsilon \ | \ \forall e' \in E \cap c \oplus B_{\epsilon}, \ d(x,e)\leq d(x,e') \} 
\end{equation} 
New connections are made between $x$ and points in the optimal visible set $\dot{ V_{x | C}}$. This one being a finite set included in $ C \cup (x \bot C)$ where $x \bot C$ is the set of orthogonal projections of $x$ on the city. 
To avoid connections too close from each other, we introduce the relative neighborhood. 
In a general way, if $P$ is a point and $E$ a points set. Then $s \in E$ is said to be in the relative neighborhood of $P$ ($ s \in RN[E || P]$) if and only if \cite{Barthelemy2009} 
\begin{equation}
\forall u \in S , d(P, s) \leq \max \{ d(u,P) , d(u, s)\}
\end{equation} 
\textit{ie} there is no point both closer to $s$ and to $P$.
All candidats to become new connections are segments from $x$ to $RN(\dot{ V_{x | C}} || x)$: $[x, RN(C || x)] $.
\subsubsection{Parameters \label{modPara}}
The tuning of those parameters will be discussed in (\ref{simulation})
\paragraph{Organization}
The global field induces minima. These minima represent points where it is the most interesting to settle. 
\\
The question is to find a parameter $P_e$ that describes whether the city is organized or not. The idea is that when a city is organized it sticks strictly to optimal settlement places and when it is purely unorganized, new settlements are added at random without any influence of the potential field. 
\\
Then a new settlement is selected by a Monte Carlo method with a number $n$ of iterations and the new point is be chosen as $X = \text{Argmin}_n \ P( X_i)$. For random cities $n$ is close to 1 and for organized one it is much higher. 
\\
Let $W$ be the area of a part of the plan that contains the current city, let $ X_1, ... , X_n$ be $n$ points on this part, uniformly and independently chosen. And let $X = \text{argmin} \ P(X_i)$. Then let 
\begin{equation}
P_e = \mathbb{P}(|X- \text{Argmin} \ P)| \leq e)
\end{equation}
represents the probability that the Monte Carlo method throws a point in a radius $e$ of a local minimum.
\\
We want to give $P_e$ as an input parameter and traduce it into an iterations number. Let $N$ be the number of local minima. If $e$ is quite small: 
\begin{equation}
P_e \approx 1 - \left(\frac{W-N.\pi.e^2}{W} \right) ^n 
\end{equation}
\begin{equation}
n \simeq \frac{\log{(1-P_e)}}{e^2}.\frac{W}{N.\pi} 
\end{equation}
$n$ is estimated roughly by noticing that a local minimum is often due to the interaction between two close streets segments: $n \sim 3.v$ since 3 is roughly the average connectivity number of intersections.
\paragraph{Connection and construction}
There are typically about four or five streets segments in $[x,RN(V_+ || x)]$ for a new center $x$. If the city shapes as a slum it would be tree like so we link the number of streets segments indeed added with the construction $ \omega \in [0,1]$ of the city. We sort segments in $[x,RN(V_+ || x)]$ by increasing length : $ (s_1, ..., s_n)$. $s_1$ is drawn with probability 1. $n' \sim \mathcal{B}(\omega, n-1) +1$ and segments $s_2,..s_{n'}$ are also added. If $\omega =1$ every admissible segment is added and if $\omega =0 $ only the shortest one.
\paragraph{Sprawling}
When constructing with a rejection radius $\lambda_0$, the city gets a typical mesh width. If at a particular urban operation a potential field with a rejection radius of $K\lambda_0$ with $K>1$ is considered then the city's inner meshes will appear as filled up with the rejection zone of this potential field and new points of interest will position outside of the city.
\\
With this observation we will consider that in a proportion $f_{ext}$ centers are added with respect to a potential of rejecting radius $K_{ext}\lambda_0$. This creates foils at the outskirt of the city and thus an extension of the city that represent for instance an industrial zone which need a large surface.
\paragraph{Refinements}
To enhance the realism of this model, some empirical parameters are added. 
\\
The length of a new street segment is bounded to $l_{\text{max}} = k_{l_{\text{max}}}\lambda_0$. This can avoid too long costly connections (possibly $l_{\text{max}} = \infty$).
\\
The windows $W$ out of where new settlements are chosen can be whether definitely set \fig{paraConstant} whether dynamically change with the overall city \fig{paraVar} and \fig{casbah}.
\\ 
Since a Monte-Carlo method is used to pick new centers, there is a very little probability that a new settlement is added in line with an existing street. If geometrically this does not have many consequences, it may strongly change the local topology. That is why, if an orthogonal projection is in a radius $c\lambda_0$ with typically $c \simeq 0.3$ of an intersection that is visible from the center then this orthogonal projection is removed. This rises the vertices degree and allows longer streets. 
\section{A few simulations \label{simulation}}
To summarize an individual simulation of the city growth we need to provide our algorithm with several parameters: 
\begin{enumerate}
\item The number of settlements: $N$
\item The organization probability $P_{e}$ 
\item The radius of the rejecting tube: $\lambda_0$ 
\item The long scale influence: $\beta$
\item The construction: $\omega$ 
\item The sprawling factor $K_{ext}$ and the sprawling probability $f_{ext}$
\end{enumerate}
Notice that only four parameters will actually shape the simulated city ($P_e$, $\beta$, $\omega$, $f_{ext}$) the others being scaling parameters and that the influence of $\beta$ won't be discussed here. 
\subsection{Simulations with constant parameters}
\begin{center}
\begin{figure}[!h]
\includegraphics[scale = 1]{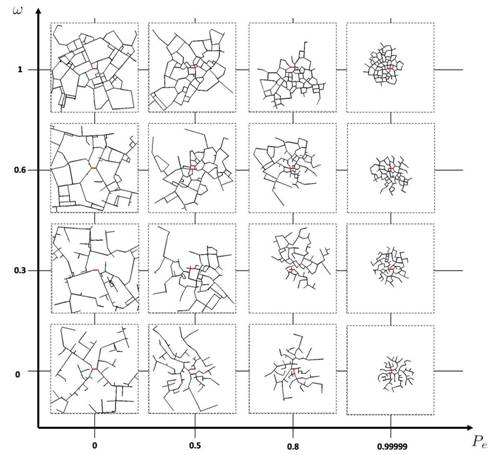}
\caption{ Simulations of the morphogenesis model with constant parameters with variation of the organization $P_e$ and the construction $\omega$. On each thumbnail, the rejecting radius is $\lambda_0= 10 m$, there is no sprawling: $k=0$, the number of settlements is $N=80$ and the available space is bound in a square with sides of $400 m$. The red and bold segment represents an initial street-segment and a scale of 20 meters.}
\label{paraConstant}
\end{figure}
\end{center}
\begin{center}
\begin{figure}[!h]
\includegraphics[scale = 1]{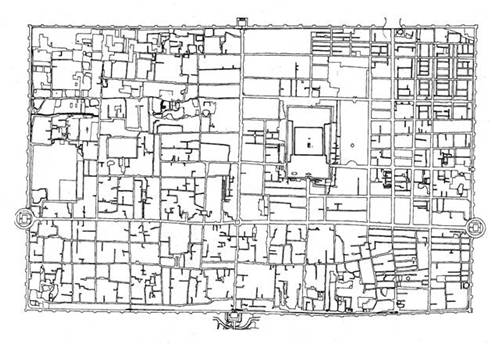}
\caption{The Chinese town of Xi'an in 1949, whose various subdivision patterns inside a regular grid recall variations in the parameters of the morphogenesis model. } 
\label{xian}
\end{figure}
\end{center}
\begin{center}
\begin{figure}[!h]
\includegraphics[scale = 1]{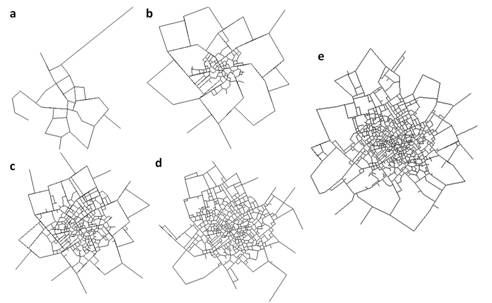}
\caption{(\textit{a, b, c, d}) : four steps in the development of the city (\textit{e} )with 600 urban operations. For this simulation, $\lambda_0=10m$, $K=10$, $K_f=0.1$, $P_e=0.8$, $\omega=0.7$ and the windows is adapted to the size of the current city. . The main phenomenon at work is the dynamics between the inner development and the extension of the city that creates two hierarchical scales.} 
\label{paraVar}
\end{figure}
\end{center}
\begin{center}
\begin{figure}[!h]
\includegraphics[scale = 1]{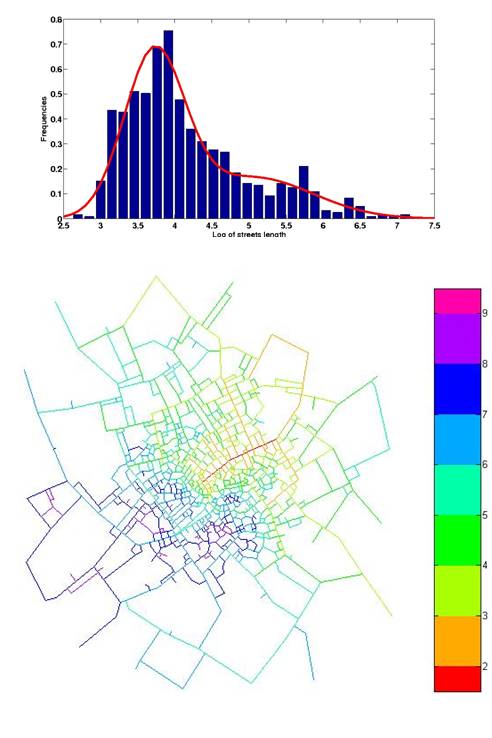}
\caption{(Color online). The distribution of the logarithm of streets length (top) and the second order topology (bottom) for the resulting city of simulation \fig{paraVar}. As for real cities (Amiens), the second order topology presents a bounded hierarchical representation of the city and the streets length is well-fitted by a mixture of log-normal random variables.} 
\label{analparaVar}
\end{figure}
\end{center}

\fig{paraConstant} shows the result of 16 simulations. The organization probability $P_e$ and the construction $\omega$ are varying jointly when the same number of operations $N=80$, the same rejecting radius $\lambda_0 = 10 m$, the same available space ( a square with an area of $1.6 \text{ km}^2$), the same initial city (a segment of length $20$ m at the center of the available space) and the same extension probability of $O$ are used. The first result is that this model is able to reproduce very different type of growth with very few "physical" parameters.
\\ 
We observe on this matrix representation that the meshedness $M_4$ (\ref{parameters}) is an increasing function of both $P_e$ and $\omega$ \tab{tabMesh}. This result has been obtained by averaging the meshedness coefficient of $30$ simulations for each couple $(P_e, \omega)$. Added to that the standard deviation of $M_4$ for each couple is of $4$ percent so that this coefficient is characteristic of the conditions of simulation. Contrary to that, the anisotropy coefficient $Ani$ is almost the same in each case (between $0.31$ and $0.46$) with a large standard deviation of $20$ percent. This $Ani$ is quite large in the absolute: for the first iterations some directions have to be arbitrary chosen, which creates favored directions. But it is the same order of weight as for the most isotropic French towns. Of course the organic ratio $r_N$ is in every case close to $1$.
\\ 
When $\omega \simeq 0$, the resulting simulations are to be compared to the Saffman-Taylor instability. It seems when $\omega \simeq 0$ that only a bounded number of ramifications are possible from the initial segment (4 on this figure) as if first created branches shielded the initial center from newer ones. When $\omega > 0$, the resulting cities are to be compared to crack patterns: their dynamics follows a logic of division / subdivision of space. 
\begin{table}
\begin{tabular}{|c|c|c|c|c|}
\hline $\Omega$ \ $P_e$
 & 0 & 0.5 & 0.8 & 0.99999 \\ 
\hline 1 & 0.37 &0.43 & 0.46 & 0.48 \\ 
\hline 0.6 & 0.26 & 0.31 & 0.33 & 0.36 \\ 
\hline 0.3 & 0.14 &0.16 &0.18 &0.20 \\ 
\hline 0 & 0 & 0 & 0 & 0 \\ 
\hline 
\end{tabular} 
\caption{\label{tabMesh}Variation of the meshedness coefficient $M_4= 2(e-v+1)/(2.v.(1-\bar{N}(2))-5)$ for the $16$ simulated cities. Each result is the averaging of $30$ simulations. The variation for each case is almost constant equal to $0.04$. $M_4$ is a increasing function of both $P_e$ and $\omega$. }
\end{table}
Interestingly, the Chinese town of Xi'an \fig{xian} has grown on a regular grid with a large mesh length with several populations that have different characteristics. The result is a gradient of meshedness coefficient, from almost $0$ in the south-west to almost $1$ in the top-right.
\fig{paraVar} presents a city evolving with $\lambda_0=10m$, $K=10$, $K_f=0.1$, $P_e=0.8$, $\omega=0.7$. The regular need of larger surface for activities such as industries, big institutions, etc. is well reproduced here. During the history, as the development of the city center progresses, it eventually absorbs the peripheral larger surfaces, to split them into smaller surfaces, with thus new larger places appearing at the new periphery. This reproduces and explains the situation of economical zones always outside at the periphery of towns. It explains as well the successive subdivision of space, that leaves so many traces, first in the log normal distribution of streets length but also in the hierarchical distributions of streets \fig{analparaVar}. For this simulation, the ratio $r_N$ is equal to $0.93$ so that the term "organic" fits. The meshedness coefficient $M_4 = 0.48$ is quite close to Amien's (between $0.41$ and $0.54$) as the anisotropy ($0.69$ to be compared to $0.71$ in the center of Amiens).
\\
\fig{analparaVar} shows that the morphogenesis reproduces the small topological radius and the log-scaling of street lengths observed in real cities (\ref{synthesis}). We have run 20 simulations with the same parameters. The root mean square distance between the length distribution and its best bi log-normal fitting range between $0.19$ and $0.26$. This compared to the result of the fittings for real French towns (less than $0.2$ when the model is good and more than $0.4$ when it is false) permits to claim that the model reproduces the street length scaling. 
\subsection{Simulations with varying parameters: the city's history}
\begin{center}
\begin{figure}[!h]
\includegraphics[scale = 0.8]{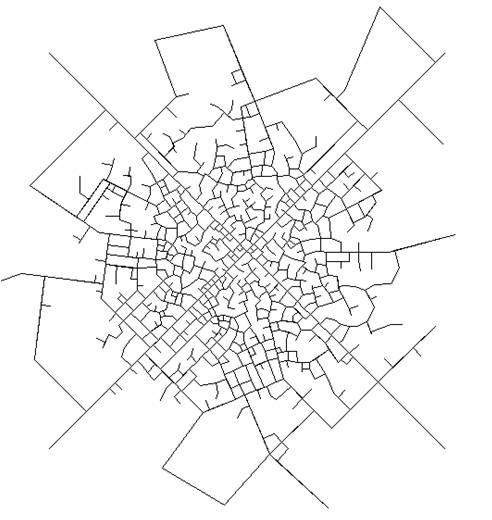}
\caption{ A city developing with varying parameters from two perpendicular long structuring elements. The historic center of the city has been built with a low construction parameter and no sprawling ($\omega = 0.2$ and $f_{ext} = 0$) to recall the tree aspect of a central souk in the Kasbah. Then parameters are changed to $\omega = 0.8$ and $f_{ext} = 0.15$ which products an industrial crown. }
\label{casbah}
\end{figure}
\end{center}

Constant parameters are not realistic to model a real city, this one being shaped by its history which is from the morphogenetic point of view a variation of input parameters.
\\ 
We represent the history of a city by a piecewise constant function $t \to (P_e, \omega, l_{\text{max}}, f_{ext}, K, \beta, \lambda_0)$. 
\\ 
For instance, the city of \fig{casbah} has been obtained by simulating at first a city with a low construction and no sprawling ($\omega = 0.2$ and $f_{ext} = 0$) and then changed to a sprawling and constructed city ($\omega = 0.8$ and $f_{ext} = 0.15$). The simulation starts with two perpendicular streets with a length $20$ times larger than $\lambda = 10 \text{m}$. These pre-existing streets are structuring elements as could be a river. This kind of variation in parameters recalls Kasbah in Morocco where the historical center of the city is a souk.
This model is the first step of a very simple model. We can see that with only one type of event (new settlement) and a few parameters a great variety of structures can be obtained. More refinements can be added. The first one is the planned creation of a highway belt as in Amiens or an enclosure using the punctual addition of the convex hull of the current city. The second possibility would be to add also the "Hausmann" effect, allowing splitting pre-existing streets with new street patterns. We could also consider distinct populations with interaction rules that build a city in the same time. These three points are going to be developed in a second version of the morphogenesis model. 
\section{Conclusion \label{conclusion}} 
We have reduced cities to the map of their street segments and shown that a lot of information can be deduced from this representation without additional data such as population dispersion, width of streets, ground specific use. For instance we can find back streets, characterize topology and shape or define a centrality. To this we have introduced the notion of geometrical and straight graph with a canonical hypergraph structure to define difference between streets segments and streets. A measure allows seeing a city's map as a "continuous graph" or likewise as an object both relational and geometrical (\ref{space}). 
\\ 
While \cite{Blanchard2009} considered a city as a pure graph and \cite{Crucitti2006} took into account its spatiality, we have explored the geometrical aspect of the city, its topology being only the skeleton that holds it up.
\\
From this point of view we have shown that despite an evident diversity on their overall shapes (anisotropy first-order topology), a few fundamental rules can explain cities' general morphogenesis (\ref{model}). The urban infrastructure differentiates to adapt to the local geography and to fulfill the constraints of lodging people while maintaining their efficient transportation.
The model developed in (\ref{model}) and illustrated in (\ref{simulation}) is based on the division / extension of space principle observed for French towns in (\ref{parameters}). It shows that structural properties of cities stand out of the local constraints and behaviors that define the dynamics of cities. For instance the log scaling of resulting street system is a global, non trivial property that validates the model as well as the small observed topological radius. Even in the organic case when there is no global and coherent plan the topological radius of the city increases slower than the size of the city. 
\\
To simulate cities' dynamics, we have uncoupled the space potential induced by the current infrastructure, the policy of connections and the freedom a new settlement (generic term to refer to a commercial infrastructure, a private individual...) ought to take on the two previous rules. 
\\ 
Further than the interest of a physical modelling of their object of study, our work can be applied by town planners and social scientists: the morphogenesis can help for semi-supervised planning and the analysis of cities allows detecting abnormalities on a map (the second-order topology representation is particularly effective). It can also be applied in engineering to test a technology that strongly depends on the urban infrastructures. 
\\
This morphogenetic model calls for a lot of outcomes: study of several potential fields, comparison to a large data basis of existing cities, following of the evolution of parameters as the city grows. The important element is that morphogenesis implements a space division and extension process. Since we used arbitrary tunnings and yet obtained realistic results, we have shown that this class of process is robust to model cities, whether it be in the centralized case or in the organic one.
\\ 
In the centralized case, the reason wants the street system to be the homotopy of a square grid to adapt to the local geography and infrastructure, which can be seen afterwards as the result of a correlated division process. 
Indeed in the organic case, the city's layout comes from the duality between the city's expansion and the graining of former large cadastres by new settlements. And from this local division process emerge some non trivial global phenomena (log scaling of streets, low topological radius). One can thus say that the division of space is a natural response of cities to fulfill their functional goals.


\begin{thebibliography}{10}%
\makeatletter
\providecommand \@ifxundefined [1]{%
 \ifx #1\undefined \expandafter \@firstoftwo
 \else \expandafter \@secondoftwo
\fi
}%
\providecommand \@ifnum [1]{%
 \ifnum #1\expandafter \@firstoftwo
 \else \expandafter \@secondoftwo
\fi
}%
\providecommand \enquote [1]{``#1''}%
\providecommand \bibnamefont  [1]{#1}%
\providecommand \bibfnamefont [1]{#1}%
\providecommand \citenamefont [1]{#1}%
\providecommand\href[0]{\@sanitize\@href}%
\providecommand\@href[1]{\endgroup\@@startlink{#1}\endgroup\@@href}%
\providecommand\@@href[1]{#1\@@endlink}%
\providecommand \@sanitize [0]{\begingroup\catcode`\&12\catcode`\#12\relax}%
\@ifxundefined \pdfoutput {\@firstoftwo}{%
 \@ifnum{\z@=\pdfoutput}{\@firstoftwo}{\@secondoftwo}%
}{%
 \providecommand\@@startlink[1]{\leavevmode}%
 \providecommand\@@endlink[0]{}%
}{%
 \providecommand\@@startlink[1]{%
  \leavevmode
  \pdfstartlink
   attr{/Border[0 0 1 ]/H/I/C[0 1 1]}%
   user{/Subtype/Link/A<</Type/Action/S/URI/URI(#1)>>}%
  \relax
 }%
 \providecommand\@@endlink[0]{\pdfendlink}%
}%
\providecommand \url  [0]{\begingroup\@sanitize \@url }%
\providecommand \@url [1]{\endgroup\@href {#1}{\urlprefix}}%
\providecommand \urlprefix [0]{URL }%
\providecommand \Eprint[0]{\href }%
\@ifxundefined \urlstyle {%
  \providecommand \doi [1]{doi:\discretionary{}{}{}#1}%
}{%
  \providecommand \doi [0]{doi:\discretionary{}{}{}\begingroup
  \urlstyle{rm}\Url }%
}%
\providecommand \doibase [0]{http://dx.doi.org/}%
\providecommand \Doi[1]{\href{\doibase#1}}%
\providecommand \bibAnnote [3]{%
  \BibitemShut{#1}%
  \begin{quotation}\noindent
    \textsc{Key:}\ #2\\\textsc{Annotation:}\ #3%
  \end{quotation}%
}%
\providecommand \bibAnnoteFile [2]{%
  \IfFileExists{#2}{\bibAnnote {#1} {#2} {\input{#2}}}{}%
}%
\providecommand \typeout [0]{\immediate \write \m@ne }%
\providecommand \selectlanguage [0]{\@gobble}%
\providecommand \bibinfo [0]{\@secondoftwo}%
\providecommand \bibfield [0]{\@secondoftwo}%
\providecommand \translation [1]{[#1]}%
\providecommand \BibitemOpen[0]{}%
\providecommand \bibitemStop [0]{}%
\providecommand \bibitemNoStop [0]{.\EOS\space}%
\providecommand \EOS [0]{\spacefactor3000\relax}%
\providecommand \BibitemShut [1]{\csname bibitem#1\endcsname}%
%</preamble>
\bibitem{Sembolonia2004}%
  \BibitemOpen
  \bibfield{author}{%
  \bibinfo {author} {\bibfnamefont{F.}~\bibnamefont{Sembolonia}}, \bibinfo
  {author} {\bibfnamefont{J.}~\bibnamefont{Assfalgb}}, \bibinfo {author}
  {\bibfnamefont{S.}~\bibnamefont{Armenib}}, \bibinfo {author}
  {\bibfnamefont{R.}~\bibnamefont{Gianassib}},\ and\ \bibinfo {author}
  {\bibfnamefont{F.}~\bibnamefont{Marsonic}},\ }%
  \bibfield{journal}{%
  \bibinfo {journal} {Computers, Environment and Urban Systems}\ }%
  \textbf{\bibinfo {volume} {28}},\ \bibinfo {pages} {45} (\bibinfo {year}
  {2004})%
  \bibAnnoteFile{NoStop}{Sembolonia2004}%
\bibitem{Frankhauser1998}%
  \BibitemOpen
  \bibfield{author}{%
  \bibinfo {author} {\bibfnamefont{P.}~\bibnamefont{Frankhauser}},\ }%
  \bibfield{journal}{%
  \bibinfo {journal} {Population}\ }%
  \textbf{\bibinfo {volume} {10}},\ \bibinfo {pages} {205} (\bibinfo {year}
  {1998})%
  \bibAnnoteFile{NoStop}{Frankhauser1998}%
\bibitem{Blanchard2009}%
  \BibitemOpen
  \bibfield{author}{%
  \bibinfo {author} {\bibfnamefont{P.}~\bibnamefont{Blanchard}}\ and\ \bibinfo
  {author} {\bibfnamefont{D.}~\bibnamefont{Volchenkov}},\ }%
  \emph{\bibinfo {title} {Mathematical analysis of urban spatial networks}}\
  (\bibinfo {publisher} {Springer Complexity},\ \bibinfo {year} {2009})%
  \bibAnnoteFile{NoStop}{Blanchard2009}%
\bibitem{Gloaguen2006}%
  \BibitemOpen
  \bibfield{author}{%
  \bibinfo {author} {\bibfnamefont{C.}~\bibnamefont{Gloaguen}}, \bibinfo
  {author} {\bibfnamefont{F.}~\bibnamefont{Fleischer}}, \bibinfo {author}
  {\bibfnamefont{H.}~\bibnamefont{Schmidt}},\ and\ \bibinfo {author}
  {\bibfnamefont{V.}~\bibnamefont{Schmidt}},\ }%
  \bibfield{journal}{%
  \bibinfo {journal} {Telecommun. Syst.}\ }%
  \textbf{\bibinfo {volume} {31}},\ \bibinfo {pages} {4} (\bibinfo {year}
  {2006})%
  \bibAnnoteFile{NoStop}{Gloaguen2006}%
\bibitem{Parish2001}%
  \BibitemOpen
  \bibfield{author}{%
  \bibinfo {author} {\bibfnamefont{Y.~I.~H.}\ \bibnamefont{Parish}}\ and\
  \bibinfo {author} {\bibfnamefont{P.}~\bibnamefont{Müller}},\ }%
  \bibfield{journal}{%
  \bibinfo {journal} {Proceedings of ACM SIGGRAPH},\ \bibinfo {pages} {301}}%
   (\bibinfo {year} {2001})%
  \bibAnnoteFile{NoStop}{Parish2001}%
\bibitem{Chen2008}%
  \BibitemOpen
  \bibfield{author}{%
  \bibinfo {author} {\bibfnamefont{G.}~\bibnamefont{Chen}}, \bibinfo {author}
  {\bibfnamefont{G.}~\bibnamefont{Esch}}, \bibinfo {author}
  {\bibfnamefont{P.}~\bibnamefont{Wonka}}, \bibinfo {author}
  {\bibfnamefont{P.}~\bibnamefont{Mü}},\ and\ \bibinfo {author}
  {\bibfnamefont{E.}~\bibnamefont{Zhang}},\ }%
  \bibfield{journal}{%
  \bibinfo {journal} {ACM Transactions on graphics}\ }%
  \textbf{\bibinfo {volume} {27}},\ \bibinfo {pages} {103} (\bibinfo {year}
  {2008})%
  \bibAnnoteFile{NoStop}{Chen2008}%
\bibitem{Porta2006}%
  \BibitemOpen
  \bibfield{author}{%
  \bibinfo {author} {\bibfnamefont{S.}~\bibnamefont{Porta}}, \bibinfo {author}
  {\bibfnamefont{P.}~\bibnamefont{Crucitti}},\ and\ \bibinfo {author}
  {\bibfnamefont{V.}~\bibnamefont{Latora}},\ }%
  \bibfield{journal}{%
  \bibinfo {journal} {Environment and Planning B: planning and design}\ }%
  \textbf{\bibinfo {volume} {33}},\ \bibinfo {pages} {705} (\bibinfo {year}
  {2006})%
  \bibAnnoteFile{NoStop}{Porta2006}%
\bibitem{Buhl2006}%
  \BibitemOpen
  \bibfield{author}{%
  \bibinfo {author} {\bibfnamefont{J.}~\bibnamefont{Buhl}}, \bibinfo {author}
  {\bibfnamefont{J.}~\bibnamefont{Gautrais}}, \bibinfo {author}
  {\bibfnamefont{N.}~\bibnamefont{Reeves}}, \bibinfo {author}
  {\bibfnamefont{R.}~\bibnamefont{Solé}}, \bibinfo {author}
  {\bibfnamefont{S.}~\bibnamefont{Valverde}}, \bibinfo {author}
  {\bibfnamefont{P.}~\bibnamefont{Kuntz}},\ and\ \bibinfo {author}
  {\bibfnamefont{G.}~\bibnamefont{Theraulaz}},\ }%
  \bibfield{journal}{%
  \bibinfo {journal} {Eur. Phys. J. B}\ }%
  \textbf{\bibinfo {volume} {49}},\ \bibinfo {pages} {513} (\bibinfo {year}
  {2006})%
  \bibAnnoteFile{NoStop}{Buhl2006}%
\bibitem{Boccalettia2006}%
  \BibitemOpen
  \bibfield{author}{%
  \bibinfo {author} {\bibfnamefont{S.}~\bibnamefont{Boccalettia}}, \bibinfo
  {author} {\bibfnamefont{V.}~\bibnamefont{Latorab}}, \bibinfo {author}
  {\bibfnamefont{Y.}~\bibnamefont{Morenod}}, \bibinfo {author}
  {\bibfnamefont{M.}~\bibnamefont{Chavezf}},\ and\ \bibinfo {author}
  {\bibfnamefont{D.-U.}\ \bibnamefont{Hwanga}},\ }%
  \bibfield{journal}{%
  \bibinfo {journal} {Physics Reports}\ }%
  \textbf{\bibinfo {volume} {424}},\ \bibinfo {pages} {175} (\bibinfo {year}
  {2006})%
  \bibAnnoteFile{NoStop}{Boccalettia2006}%
\bibitem{Gross2004}%
  \BibitemOpen
  \bibfield{author}{%
  \bibinfo {author} {\bibfnamefont{J.~L.}\ \bibnamefont{Gross}}\ and\ \bibinfo
  {author} {\bibfnamefont{J.}~\bibnamefont{Yellen}},\ }%
  \emph{\bibinfo {title} {Handbook of graph theory}}\ (\bibinfo {publisher}
  {CRC Press},\ \bibinfo {year} {2003})%
  \bibAnnoteFile{NoStop}{Gross2004}%
\bibitem{Crucitti2006}%
  \BibitemOpen
  \bibfield{author}{%
  \bibinfo {author} {\bibfnamefont{P.}~\bibnamefont{Crucitti}}, \bibinfo
  {author} {\bibfnamefont{V.}~\bibnamefont{Latora}},\ and\ \bibinfo {author}
  {\bibfnamefont{S.}~\bibnamefont{Porta}},\ }%
  \bibfield{journal}{%
  \bibinfo {journal} {Physical Review E}\ }%
  \textbf{\bibinfo {volume} {73}},\ \bibinfo {pages} {036125} (\bibinfo {year}
  {2006})%
  \bibAnnoteFile{NoStop}{Crucitti2006}%
\bibitem{Jiang2004}%
  \BibitemOpen
  \bibfield{author}{%
  \bibinfo {author} {\bibfnamefont{B.}~\bibnamefont{Jiang}}\ and\ \bibinfo
  {author} {\bibfnamefont{C.}~\bibnamefont{Claramunt}},\ }%
  \bibfield{journal}{%
  \bibinfo {journal} {Environment and Planning B : Planning and design}\ }%
  \textbf{\bibinfo {volume} {31}},\ \bibinfo {pages} {151} (\bibinfo {year}
  {2004})%
  \bibAnnoteFile{NoStop}{Jiang2004}%
\bibitem{Cardillo2006}%
  \BibitemOpen
  \bibfield{author}{%
  \bibinfo {author} {\bibfnamefont{A.}~\bibnamefont{Cardillo}}, \bibinfo
  {author} {\bibfnamefont{S.}~\bibnamefont{Scellato}}, \bibinfo {author}
  {\bibfnamefont{V.}~\bibnamefont{Latora}},\ and\ \bibinfo {author}
  {\bibfnamefont{S.}~\bibnamefont{Porta}},\ }%
  \bibfield{journal}{%
  \bibinfo {journal} {Physical Review E}\ }%
  \textbf{\bibinfo {volume} {73}},\ \bibinfo {pages} {066107} (\bibinfo {year}
  {2006})%
  \bibAnnoteFile{NoStop}{Cardillo2006}%
\bibitem{Barthelemy2009}%
  \BibitemOpen
  \bibfield{author}{%
  \bibinfo {author} {\bibfnamefont{M.}~\bibnamefont{Barthélemy}}\ and\ \bibinfo
  {author} {\bibfnamefont{A.}~\bibnamefont{Flammini}},\ }%
  \bibfield{journal}{%
  \bibinfo {journal} {Networks and spatial economics}\ }%
  \textbf{\bibinfo {volume} {9}},\ \bibinfo {pages} {401} (\bibinfo {year}
  {2009})%
  \bibAnnoteFile{NoStop}{Barthelemy2009}%
\end{thebibliography}
\end{document}